\documentclass[11pt,preprint]{aastex}

\def\dg{{\rm o}}
\def\be{\begin{equation}}
\def\ee{\end{equation}}
\def\Sgr{Sgr~A$^*$~}

\newbox\grsign \setbox\grsign=\hbox{$>$} \newdimen\grdimen \grdimen=\ht\grsign
\newbox\simlessbox \newbox\simgreatbox \newbox\simpropbox
\setbox\simgreatbox=\hbox{\raise.5ex\hbox{$>$}\llap
     {\lower.5ex\hbox{$\sim$}}}\ht1=\grdimen\dp1=0pt
\setbox\simlessbox=\hbox{\raise.5ex\hbox{$<$}\llap
     {\lower.5ex\hbox{$\sim$}}}\ht2=\grdimen\dp2=0pt
\setbox\simpropbox=\hbox{\raise.5ex\hbox{$\propto$}\llap
     {\lower.5ex\hbox{$\sim$}}}\ht2=\grdimen\dp2=0pt

\begin{document}

\title{
Stellar disk in the galactic center --- a remnant
of a dense accretion disk?} 

\author{Yuri Levin\altaffilmark{1}, Andrei M. Beloborodov\altaffilmark{1,2,3}} 

\altaffiltext{1}{Canadian Institute for Theoretical Astrophysics,
60 St. George Street, Toronto, ON M5S 3H8, Canada}

\altaffiltext{2}{Physics Department, Columbia University, 538  West 120th 
Street New York, NY 10027}

\altaffiltext{3}{Astro-Space Center of Lebedev Physical 
Institute, Profsojuznaja 84/32, Moscow 117810, Russia}

\begin{abstract}

Observations of the galactic center
revealed a population of young massive stars
within $0.4$~pc from \Sgr --- the presumed
location of a supermassive black hole. The origin
of these stars is a puzzle as their formation {\it in citu}
should be suppressed by the black hole's tidal field.
We find that out of 13 stars whose 3-dimensional
velocities have been  measured by Genzel et.~al.~(2000),
10 lie in a thin disk. The half-opening angle of the disk is
consistent with zero within the measurement errors, and does not
exceed 10 degrees. We propose
that a recent burst of star formation has occurred in a dense
gaseous disk around \Sgr. Such a disk is no longer present
because, most likely, it has been accreted by the central black hole.

The three-dimensional orbit of S2, the young star closest to
\Sgr, has been recently mapped out with high precision.
It is inclined to the stellar disk by 75 degrees. We find
that the orbit should undergo Lense-Thirring precession with
the period of $\sim (5/a)\times 10^6$ years, where $a<1$ is the
dimensionless spin of the black hole. Therefore it is possible
that originally S2 orbit lay in the disk plane. If so, we can
constrain the black hole spin $a$ be greater than
$0.2\times (t_{\rm S2}/5\times 10^6\hbox{yr})$, where
$t_{\rm S2}$ is the age of S2.

\end{abstract}



\section{Introduction}

The dusty center of our galaxy harbors a compact massive object
\Sgr, most likely a black hole (Genzel et.~al.~1997, Ghez et.~al.~1998).
The galactic center is also host to a numerous
population of young stars (see Genzel et.~al.~2000, hereafter
G00, and references therein). Such a combination is
a puzzle as the tidal field of \Sgr should suppress star
formation in its vicinity (Phinney 1989). 
Interestingly, G00 have found that most young stars move clockwise on 
the sky around the putative black hole.

In this Letter we analyze and interpret the three-dimensional (3D) 
velocity data of G00. In the next section, we show the available sample 
of 13 stars and find that 10 of them revolve in a thin disk.
We discuss this result in \S~3 and 
propose that a burst of star formation occurred recently within a dense 
gaseous disk. In \S~4, we consider the orbit of S2, the young star
closest to \Sgr (Sch\"odel et al. 2002; Ghez et al. 2003).
Its orbital plane is inclined by $75^\dg$ to the 
stellar disk. We show that
the orbital plane of S2 is turning around the black-hole spin
axis due to the Lense-Thirring precession, and thus it is 
possible that S2 was born in the plane of the stellar disk. 
Within this hypothesis, we are able to constrain
the magnitude of the black-hole spin.

\section{The young stellar disk}

For 13 young stars with strong He-emission lines, G00 were able 
to measure 3D velocities. Eleven of them were found to move clockwise 
on the sky. We find that 10 velocity vectors of the clockwise-moving 
stars from this sample lie in one plane within the measurement errors. 

Figure~1 shows the full available sample of 13 vectors
in two projections: (a) as viewed by an observer on Earth and
(b) as viewed by an observer located in the found plane.
One can see that ten  vectors lie in the plane within at most
$1.6\sigma$. The sample is clearly inconsistent with
a stellar population of randomly oriented orbits.
Indeed, what would be the probability to have
10 out of 13 random vectors clustering so tightly near
a single plane? A Monte-Carlo simulation
demonstrates that this probability is $P\approx 10^{-2}$.
If we limit our sample to stars 1---11 that move
clockwise on the sky, then $P\approx 10^{-3}$.

The pattern of velocities observed in Figure~1b
suggests that the stars are divided into two groups.
(1) Stars 1---10 with nearly co-planar velocities
belong to a thin disk; note that they all move clockwise on the sky.
(2) Stars 11---13 with velocities far from the disk
plane ($>3\sigma$); they have a random sense of rotation
on the sky (stars 12 and 13 move counter-clockwise).

We fitted vectors 1---10 by one plane using the $\chi^2$ test.
The reduced $\chi^2$ is defined by
\begin{equation}
\chi^2={1\over N-1}\Sigma_{i=1}^N {({\bf n}\cdot {\bf v}_i)^2
\over (n_x\sigma_{xi})^2+(n_y\sigma_{yi})^2+(n_z\sigma_{zi})^2},
\label{chi2}
\end{equation}
where $N$ is the number of stars,
${\bf n}=(n_x, n_y, n_z)$ is a unit vector normal to the plane,
${\bf v}_i$ is the velocity vector of the $i$th star, and
$(\sigma_{xi}, \sigma_{yi}, \sigma_{zi})$ are the
measurement errors of the components of ${\bf v}_i$. 
The best fit has the reduced $\chi^2=0.67$.
If we added the 11th star then $\chi^2$ would jump to 3.2.
Unit vector ${\bf n}$ normal to
the best-fit plane is $(0.486, 0.725, -0.487)$ in
conventional galactic coordinates: $x$ and $y$ are the coordinates
on the sky (right ascension and declination, respectively) and
$z$ axis is directed along the line of sight away from us.
The $\chi^2$ test also shows that the uncertainty in ${\bf n}$
is small: $\chi^2$ increases by a factor of two if one rocks
$\bf{n}$ through $9^\dg$.

Although the data is consistent with stars 1---10 having orbits
exactly in the best-fit plane, the measurement errors allow modest
orbital inclinations relative to this plane, $\theta_i\neq 0$
($i=1,...,10$), i.e. the stellar disk can have a modest openning angle. 
We now evaluate the upper limit on this angle.
Suppose the ten stars are drawn from a rotating stellar system
with orbital planes pivoting around the best-fit plane with a
dispersion $\Delta\theta$.
To check the consistency of a given $\Delta\theta$ with the data we use
the Monte-Carlo technique. We simulate a random realization of ten 
orbital planes $i=1,...,10$ drawn from the model population, and choose 
in each plane a vector ${\bf v}_i$ with the absolute value equal to the
corresponding real $v_i$ and a random direction\footnote{
To mimic the measurement procedure,
we also superimpose random Gaussian errors $\delta{\bf v}_i$ on the
generated vectors. Their mean deviations
$(\sigma_{xi},\sigma_{yi},\sigma_{zi})$ are taken equal to those
of the real measurements of G00.}.
Then, using equation~(1), we fit ${\bf v}_i$ by one plane.
With many realizations of the artificial sample,
we find the probability that $\chi^2\leq 0.67$.
In other words, we find the probability that the artificial sample
allows as good co-planar fit as the one we found for the real data.
This probability $P(\Delta\theta)$ is a measure of the data consistency 
with assumed $\Delta\theta$. The results are shown in Figure~2.
We find $P\approx 10^{-2}$ for $\Delta\theta=20^\dg$
and $P\approx 10^{-1}$ for $\Delta\theta=10^\dg$.
Thus, the half-opening angle of the stellar disk
does not exceed $\sim 10^\dg$.

\section{The interpretation: disrupted young cluster versus
 remnant of an accretion disk}

The existence of the young stellar disk provides a new insight into
the  puzzle of star formation in the galactic center. Normal star
formation via fragmentation of a molecular cloud is impossible there,
since the cloud would be ripped apart by the tidal forces it would 
encounter so close to the black hole (Phinney 1989). One solution to 
this problem was proposed by Gerhard (2001). In his picture, 
a stellar cluster was born $\sim 10$~pc away from \Sgr,
and then spiraled in due to gravitational interaction
with ambient stars. As it spiraled in, the cluster was
stripped and finally its core was disrupted by the tidal
field of \Sgr.
The young stars left from the disintegrated cluster would continue
to move in orbital planes close  to that of the parent cluster,
within an angle $\Delta\theta\sim(M_c/M)^{1/3}$, where
$M_c$ is mass of the cluster core and $M$ is the black-hole mass.
This is consistent with the observed thickness of
the stellar disk if $M_c< 5\times 10^{-3}M\approx
2\times 10^4M_\odot$, which also agrees with Gerhard's estimate
for $M_c$. However, this scenario has a problem:
the cluster should be disrupted
relatively far from \Sgr, at about 0.4~pc (Gerhard 2001).
In contrast, many of the young stars are located within 0.1~pc,
thus their origin remains unexplained.

We propose  that the  stars were born in a dense gaseous
disk that existed around \Sgr in the past.
Such a disk can form when an infalling molecular
cloud is tidally disrupted by the central black hole. Alternatively,
the disk can grow gradually by accumulating gas from the ambient
medium. In either case, the disk orientation is
determined by the gas angular momentum.  Accumulation
of gas in one plane leads to a significant increase of its density.
When the
 density exceeds $M/(2\pi R^3)$ at a radius $R$,  
the disk become unstable with respect to its own gravity,
and its fragments collapse (Paczynski 1978). The clumping instability
can lead to formation of stars, and this process is
believed to occur  in active nuclei of other galaxies
(Kolykhalov and Sunyaev 1980, Shlosman and Begelman 1987,
Collin and Zahn 1999).
Detailed models  of self-gravitating disks show that the fragmentation 
takes place at radii $R=(10^4-10^5)GM/c^2$ (Goodman 2003),
which is consistent with the proximity of the young stellar
population to Sgr~A$^*$. The dense accretion disk can also trap old 
stars from initially inclined orbits (Artymowicz, Lin, and Wampler 1993). 
The trapped stars would then accrete the disk material and grow to large 
masses, so that now they would appear as young stars.

No dense gaseous disk is currently observed around \Sgr, which
implies that it is a transient phenomenon. After the onset of the
clumping instability and the burst of star formation, the disk
can be drained quickly: partly it is blown away by the stellar
winds and partly it is accreted onto the central black hole.
A major remnant of the past activity near \Sgr are the young stars.
The estimated  ages of stars indicate that the burst of accretion
took place $3-9$ millions of years ago.

The stellar disk is immersed in the ambient spherical cluster of old stars,
and   is being stirred  by their gravitational fields.
With time, the disk will get thicker and, in approximately
$30$ million years, it will  become indistinguishable from
the spherical cluster (see G00 for the estimate of the cluster relaxation
time). The present age of the young stars is significantly
smaller and the disk can be well preserved.
The fact that a fraction of the young stars do
not belong to the disk remains to be explained.
In particular, 3 stars with measured 3D velocities
are out of the disk plane, and 2 of them are on counter-rotating orbits.

\section{The Lense-Thirring precession of close stars}

The available sample used in our work does not include a comparable 
number of stars that are even closer to \Sgr but whose 3D velocities 
remain unmeasured. In the accretion-disk scenario, they could have
migrated inward as a result of tidal interaction with the dense
disk (Ward 1986, Gould and Rix 2000).
The lack of 3D velocity measurements of close stars does not allow us
to check whether their orbits are in the disk plane.
Recently, it became possible to map out the
3-dimensional orbit of star $S2$, which is  closest to the black
hole (Sch\"odel et.~al.~2002, Ghez et.~al.~2003).
 Its sense of orbital rotation is similar to that of
the stellar disk, however, the orbital plane is different.
There is no inconsistency here: S2 is so close to \Sgr that
the black-hole spin can make the orbital plane of S2 turn
slowly around the spin axis.
This is a general-relativistic effect known as Lense-Thirring
precession (Misner, Thorne, and Wheeler 1973).
Using the measured orbit eccentricity $e=0.87$ and semimajor
axis $r_0=4.6\times10^{-3}$~pc, we find the period of precession,
\begin{equation}
P_{\rm LT}=5.76\times 10^6a^{-1}\left({3\times
   10^6M_{\odot}\over M_{}}\right)^2
           \left({1-e^2\over 1-0.87^2}\right)^{3/2}
   \left({r_0\over 4.6\times 10^{-3}\hbox{pc}}\right)^3\,\hbox{yr},
 \label{lensethirring}
\end{equation}
where $a$ is the dimesionless spin of the black hole, with $a=1$
corresponding to a maximal spin.
Thus, even though S2 orbit is not presently  in the plane of the
stellar disk, it could have been there a few million years ago.
The current
angle between the disk and S2 orbital plane is about $75^\dg$,
and the disk origin of S2 would require
\begin{equation}
  a>0.21\left({3\times10^6M_{\odot}\over M_{\rm BH}}\right)^2
         \left({5.76\times10^6\hbox{yr}\over t_{\rm S2}}\right),
\label{amin}
\end{equation}
where $t_{\rm S2}$ is the age of $S2$.

Once the orbits of other close stars are mapped out,
 the hypothesis that they
were born in the stellar disk and later
precessed out of the disk plane can be tested.
Just one additional orbit will be sufficient to
determine the direction of the black-hole spin,
and two additional orbits would allow one to confirm or
rule out the precession hypothesis.

\acknowledgments
We are grateful to Reinhard Genzel for the update on the stellar velocities,
and to Norm Murray, Marten van Kerkwijk, and Chris Matzner for discussions.
We thank Sarah Levin for her advise on the prose.
This work was supported by NSERC and RFBR grant 00-02-16135.


\begin{figure}
\begin{center}
\plotone{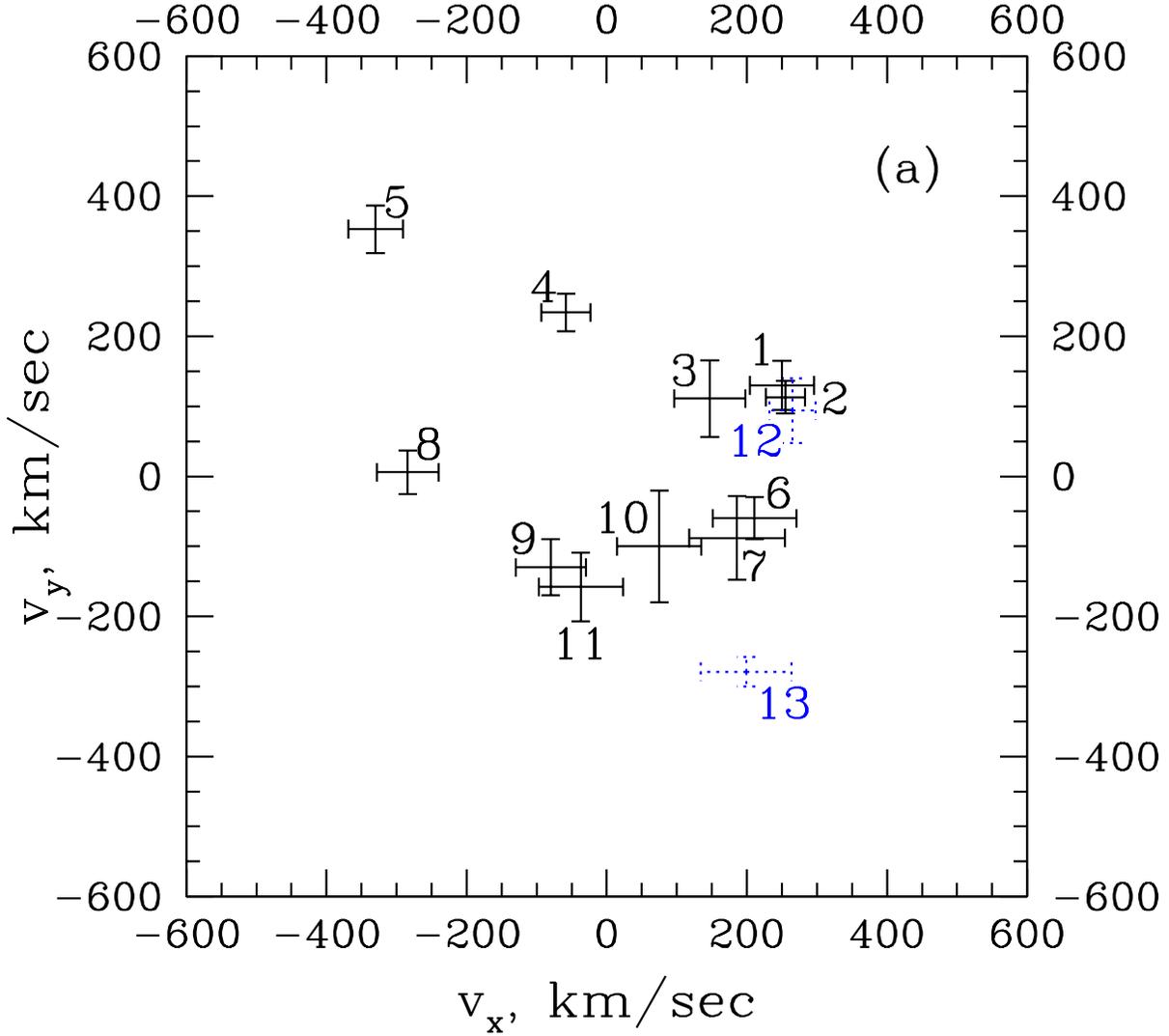}
\caption{ Velocities of the 13 stars.
(a) Projected on the plane of the sky. (b) As viewed by an observer located
in the found disk plane. The plane is shown by the heavy horizontal line.
The data are taken from Table~I of G00.
The names of stars marked by 1,...,13 are: 16SW, 16SE1, 16SE2, 16CC,
16C, 33E, 29N, 29NE1, 34W, 7W, W10, 16NW, 16NE.
Stars 1---11 move clockwise on the sky,
and 12, 13 move counter-clockwise.
$v_n$ and $v_q$ are velocity projections on the disk normal
{\bf n} and unit vector ${\bf q}$ which lies in the disk;
${\bf n}=(0.486, 0.725, -0.487)$ and ${\bf q}=(-0.271,-0.407,-0.877)$
in the galactic coordinates $(x,y,z)$.
}
\end{center}
\end{figure}

\begin{figure}
\begin{center}
\plotone{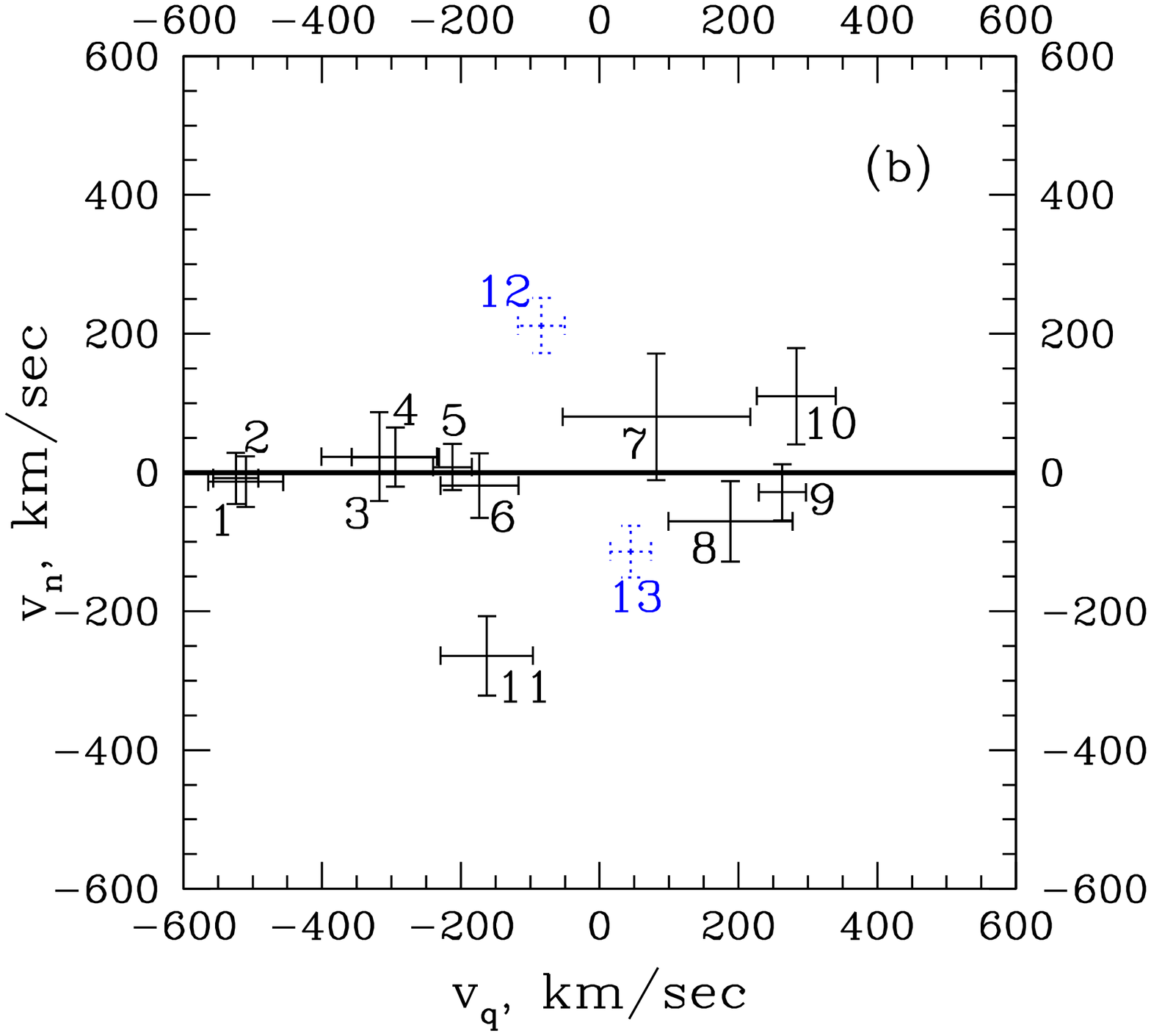}
\end{center}
\end{figure}

\begin{figure}
\begin{center}
\plotone{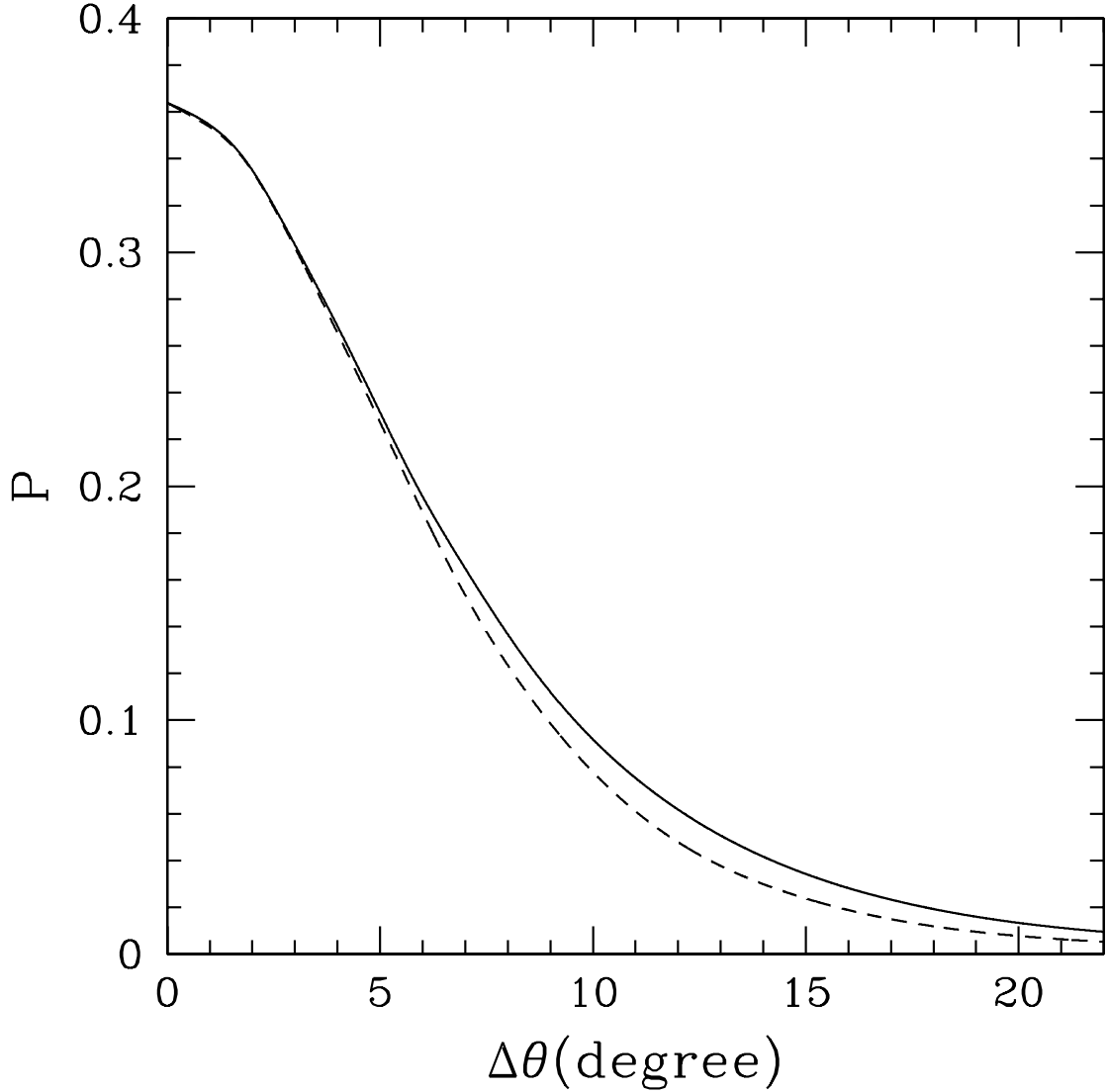}
\caption{ Constraint on the disk opening angle.
We test the hypothesis that stars 1---10 are randomly drawn from a disk 
with a dispersion of orbital inclinations $\Delta\theta$.
The plot shows the probability $P$ that 10 randomly drawn stellar 
velocities can be fitted by one plane with $\chi^2\leq 0.67$.
Solid line shows $P(\Delta\theta)$ for a Gaussian distribution
of orbital inclinations $\theta$, and dashed line shows the
case of a homogeneous distribution.
}
\end{center}
\end{figure}

\end{document}